# Formation of wrinkles on graphene induced by nanoparticles: atomic force microscopy study

B. Pacakova[1,2], J. Vejpravova[1*], A. Repko[2,3], A. Mantlikova[1], M. Kalbac[4]

[1]Institute of Physics of the ASCR, v.v.i., Na Slovance 2, 182 21 Prague 8, Czech Republic.

[2]Charles University in Prague, Faculty of Mathematics and Physics, Ke Karlovu 3, 121 16 Prague 2, Czech Republic.

[3]Charles University in Prague, Faculty of Science, Albertov 6, 128 43 Prague 2, Czech Republic.

[4]J. Heyrovsky Institute of Physical Chemistry of the ASCR, v.v.i., Dolejskova 3, 182 23 Prague 8, Czech Republic.

**Abstract:**

Wrinkles in monolayer graphene (GN) affect the GN electronic and transport properties. Defined network of wrinkles can be reached by placing the GN on the substrate decorated with the nanoparticles (NPs). In order to explain mechanism behind the topographically induced changes of the GN electronic structure and to correlate it with the wrinkling, correct description of the GN morphology is of high demand. We demonstrate here how to determine from the atomic force microscopy (AFM) images whether the GN is in the contact with the substrate or it is delaminated, how is the wrinkling of the layer connected with the density of the NPs and whether there is the preferential orientation of wrinkles or not. Also the relevance of detection of the NPs hidden beneath the GN layer is discussed. The study was done on the samples of the GN on the top of $SiO_2$/Si substrate decorated with the metal-oxide NPs of nominal diameter of the 6 and 10 nm. NP density varied in a range of 20 – 470 NPs/$\mu m^2$. The projected area of wrinkles was increasing linearly with the increasing NP density, independently on the NP size.

*e-mail: vejpravo@fzu.cz, phone: +420 266 052 325

1. **Introduction**

Creation of wrinkles is a property of any constrained thin sheet [1–5], self-similar hierarchy of wrinkling is established both for a thin fabric as well as suspended graphene (GN) [1]. Moreover, GN itself is intrinsically non-flat [6,7], because ideal 2D flat crystals are unstable to thermal fluctuations [6–8]. If the GN is transferred to the substrate, the wrinkles naturally created on as-grown layer are either released or preserved, depending on the morphology of underlying substrate and topography of supported GN which differs from intrinsic corrugations in the free standing layer [2,7,9–11].

In order to develop the GN with distinct level of wrinkles (means to influence density of wrinkles, percentage of the GN layer covered with wrinkles, height of wrinkles, mean orientation of wrinkles and others), two approaches were already presented in the literature: transfer of the GN on elastic substrate [12] or on decorated substrates, covered either with the 50 nm $SiO_2$ NPs [13,14] or nanopillars of varying height and curvature of the top of pillars in a range of tens-hundreds of nm [15–17].

Wrinkling phenomena in GN appeared to be interesting for several reasons. At first, electronic and transport properties known for non-perturbed GN [18,19] are affected by presence of corrugations [20–24]. The scattering of the charge carriers is influenced by any topographical inhomogeneity in the GN layer [25,26] and is manifested for example as decreased and/or fluctuating carrier mobility and charge density [23,27,28] or creation of electron-hole puddles [20,25,29,30]. Formation of wrinkles, more specifically delamination of the GN from substrate is connected with the doping of the GN layer from the substrate [14,31,32]. Delaminated parts of the GN are the wrinkles themselves, but for the supported GN, smooth parts of the GN among the wrinkles can be also elevated above the substrate. Hence doping can be affected in desired way when the well-defined part of the GN is wrinkled and delaminated [14,32].

Finally, creation of wrinkles is connected with the introduction of the shear strain into the GN layer [4,14,33,34] and hence induced change of the electronic properties of GN [15,24,35–41]. It was reported that for specific strain symmetries (three-axial, $C_{3v}$), the local potential centers are formed in the GN layers [42,43]. Strain can be induced both into the flat part of the GN (wrinkles propagating in three different directions can induce such three-axial strain [36,37]) and GN parts directly capping the supporting nanoparticles (NPs) [13]. The consequence of creation of local potential center is the postulation of existence of the so-called "pseudofields" [42,43]. As far the necessity of existence of short-range potential for modification of the magnetoresistance of graphene was reported [44], GN supported by the NPs also seems to be good candidate for examination of this phenomena. Fundamental understanding of the topographical properties of wrinkled GN and their relation to electronic structure is important to assign the underlying mechanism.

In our work, we systematically studied morphology and wrinkling hierarchy of the GN layer covering the substrate decorated with the NPs. We prepared eleven samples of the CVD-grown GN transferred on the $Si/SiO_2$ substrates decorated with the 6 and 10 nm metal-oxide NPs with surface density within a range of 20 – 470 NPs/$\mu m^2$ and demonstrated that defined wrinkling of the GN can be reached by changing the NP surface density.

We present here the statistical analysis and data processing of the Atomic Force Microscopy (AFM) images of wrinkled GN on decorated substrates and specify relation between creation of wrinkles and the density of the NPs on the substrate. It is shown that delamination of the GN from substrate and preferential orientation of wrinkles can be determined just with the use of the AFM. Relevance of detection of the NPs beneath the GN is also discussed.

## 2. Material and methods

### 2.1. Preparation of the GN@NPs samples

The samples of GN supported by the NPs were prepared by spin-coating of the NP dispersions with varying concentration on the substrates and subsequent transfer [45] of the CVD grown GN layer [46,47] on the top. Two batches of the samples were prepared, first batch with use of the $CoFe_2O_4$ NPs with mean diameter ~ 6 nm [48] (the GN_1-GN_6 samples), second batch with the γ-$Fe_2O_3$ NPs with mean diameter of 10 nm [49] (GN_7-GN_11 samples). Details of the preparation of the first sample batch and parameters of the spin-coating of the NPs can be found in paper by Vejpravova et al. [32]. The first batch of the samples was exposed to 15 min drying at 300 °C in air prior to the GN transfer, second batch was dried after transfer of GN.

### 2.2. Atomic force microscopy (AFM)

The AFM images were captured at ambient conditions with the Veeco Multimode V microscope equipped with the JV scanner, with the resolution of 1024 lines. Fresh RFESP probe ($k = 3$ N/m, $f_0 = 75$ kHz, nominal tip radius = 8 nm) by Bruker was used for each sample, preserving the wear of the tip comparable. All images were captured in the standard tapping mode in order to minimize force acting on the GN layer. The scan rate was 0.8 Hz. The amplitude setpoint was chosen as near to the free amplitude as was possible. Several images of decorated substrates as well as the GN layer were captured, with the maximum image size of 25 μm². The captured images were processed afterwards in the Gwyddion software [50] by 1[st] or 2[nd] order flattening and scars correction. The root-mean square roughness [51,52], $σ_p$ of the bare substrates, decorated substrates, smooth and wrinkled parts of the GN layer were determined.

The objects on the surface (both the NPs on substrate and wrinkles on the GN layer) were located by height and slope threshold masking method [50] [53,54], setting the threshold height and slope for each image individually. Masking of selected objects/regions on substrate enables to evaluate properties of masked and unmasked parts of the image separately.

Grain analysis of masked images (Gwyddion) was used to determine the number of the NPs in the image of decorated substrate of defined area. Then the surface NP density, $\rho_{NP}$ was calculated. The histograms of equivalent disk radius, $r_{eq}$ were created for localized NPs, distributions of the $r_{eq}$ values were refined with the log-normal distribution function. Mean radius of the NPs for each individual sample was determined. The ($x,y$) coordinates of each single localized NP were used to calculate the mean interparticle distance, $d_{real}^{mean}$ by the triangulation method [55]. The triangulation method was implemented in the Matlab code for simulations of the random distribution of the NPs in defined area (see section 2.3.).

For the images of the GN layers, the value of the masked projected area attributed to wrinkles, $A_{wr}$ was determined for each image.

In further analysis, we used basic property of the AFM image, which is an array of pixels with the ($x,y,z$) coordinates. It is easy to collect the $z$ coordinates for whole image and to create the height histogram ($z$-histogram) for whole image as well as masked and unmasked regions separately. $z$-histograms were created both for decorated substrates and the GN layer. Images of the GN layer used for creation of $z$-histograms were chosen to always contain at least small rupture in the GN in order to "see" the substrate beneath and set the zero height of the image as the minimum height of the substrate. $z$-histograms of whole images and images with excluded masked regions were constructed in order to specify the heights attributed to the objects (NPs or wrinkles) and the flat parts of images (substrate, smooth parts of the GN).

The height-height autocorrelation functions (HHCF) both of decorated substrates as well as compact GN layer were generated for each individual sample, obtained curves were compared with the Gaussian HHCF [52,56].

Two-dimensional fast Fourier transformation (2D FFT) was applied to images of compact GN layers in order to determine the sharp edges at the image and hence identify local symmetry of the wrinkles and their preferential orientation.

In order to locate the NPs beneath the GN, the analysis of phase images was done. Areas of the images with elevated phase values were attributed to the NPs. Particle search was done by the threshold masking method. Details are summarized in the Supplementary information file.

**2.3. Simulations of model decorated substrates**

Random distribution of the NPs on substrates was simulated in the MATLAB network. 2D array of randomly distributed NPs placed into the area of 5x5 $\mu m^2$ was simulated to represent each individual real sample. Number of the NPs in the array was calculated using the $\rho_{NP}$ value of each individual real sample. Then the mean interparticle distance for simulated distribution of the NPs, $d_{mean}^{sim}$ was determined by the triangulation method [55] from the coordinates of individual simulated NPs, ($x_{sim}$, $y_{sim}$).

3. **Results and discussion**

We captured multiple AFM topography images of decorated substrates and the whole GN@NPs samples. All samples displayed typical wrinkles (see Figure 1, Figure 2c and Figure S1-S5 in the Supplementary information file).

The analysis of the AFM images of substrates, decorated substrates and whole GN@NPs samples revealed that the $\sigma_p$ of the GN layer placed on decorated substrate is significantly lower than is the $\sigma_p$ of decorated substrates, and in the range of the error corresponds with the $\sigma_p$ of the bare substrate (Figure 1c).

The substrates decorated with the NPs as well as wrinkled GN layers were subjected to threshold masking procedure to locate individual NPs and wrinkles. Distributions of the NP radius (diameters) determined from the AFM images, so-called $r_{eq}$-histograms revealed that distribution of the $r_{eq}$ is the log-normal, which is typical and qualitatively agrees with distribution of the NP size in initial NP solutions for NPs prepared by decomposition of organic precursor [48,49,57] (example Figure 2a and S7 in the Supplementary information file). It was demonstrated that the NPs created aggregates in specific cases (the GN_2 and GN_5 samples).

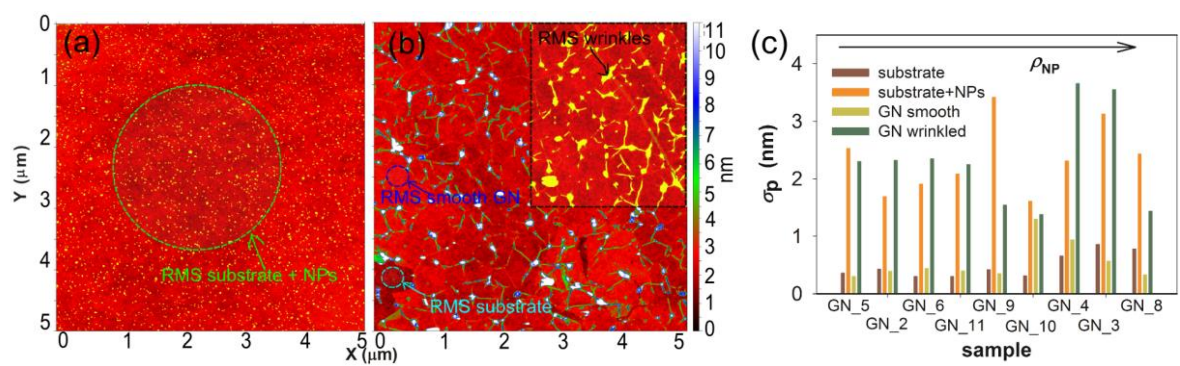

Figure 1. (a, b) Illustration of the specific areas at the AFM height images from which the root mean square (RMS) roughness, $\sigma_p$ is determined: namely $\sigma_p$ of substrate, substrate decorated with the NPs (substrate + NPs), smooth GN and GN wrinkles. The area with masked wrinkles (yellow mask) is bordered with the dashed lines (b). (c) $\sigma_p$ for individual samples and their parts, determined from the 25 μm² area.

The $\rho_{NP}$ for individual samples was obtained simply evaluating the number of the NPs in the imaged area (Table 1). Figure 2c visualizes the substrates with the low (GN_2) and high (GN_3) values of the $\rho_{NP}$, images of the rest of the samples are depicted in the Supplementary information file (Figure S1-5).

The real interparticle distances, $d_{mean}^{real}$ were determined by the triangulation method, using determined ($x,y$) coordinates of individual NPs (values are listed in Table 1). Then the simulation of random distribution of the NPs representing the real samples was done using the $\rho_{NP}$ of individual samples as an input, the $d_{mean}^{sim}$ was determined (Supplementary information file, Table S1). Comparison of $d_{mean}^{real}$ and $d_{mean}^{sim}$ values showed that values are comparable, hence the interparticle distances of the NPs in real samples correspond to those of random distribution of objects on the substrate.

**Table 1.** NP density, $\rho_{NP}$ determined for decorated substrates, mean real interparticle distances, $d_{mean}^{real}$, distance between smooth flat parts of (see Figure 2b) graphene and substrate, $d_{GN-substrate}$ and interpretation (whether the examined part of the GN is in the contact with the SiO$_2$/Si substrate or it is delaminated).

| sample | $\rho_{NP}$ (NPs/μm$^2$) | $d_{mean}^{real}$ (nm) | $d_{GN-substrate}$ (nm) | interpretation |
|---|---|---|---|---|
| **GN_5** | 20 ± 3 | 267.7 ± 1.8 | 4.6 ± 0.2 | Delamination |
| **GN_2** | 77 ± 5 | 139.6 ± 4.7 | 1.0 ± 0.1 | Contact |
| **GN_6** | 134 ± 8 | 104.0 ± 3.9 | 3.6 ± 0.2 | Delamination |
| **GN_11** | 150 ± 9 | 106.3 ± 3.8 | 0.4 ± 0.2 | Contact |
| **GN_9** | 156 ± 9 | 96.7 ± 3.7 | 1.1 ± 0.2 | Contact |
| **GN_10** | 168 ± 10 | 92.9 ± 3.6 | 3.7 ± 0.3 | Delamination |
| **GN_4** | 438 ± 44 | 54.5 ± 4.4 | 2.0 ± 0.1 | Contact |
| **GN_3** | 454 ± 45 | 54.2 ± 4.3 | 1.6 ± 0.1 | Contact |
| **GN_8** | 470 ± 47 | 58.0 ± 4.3 | 3.0 ± 0.2 | Delamination |

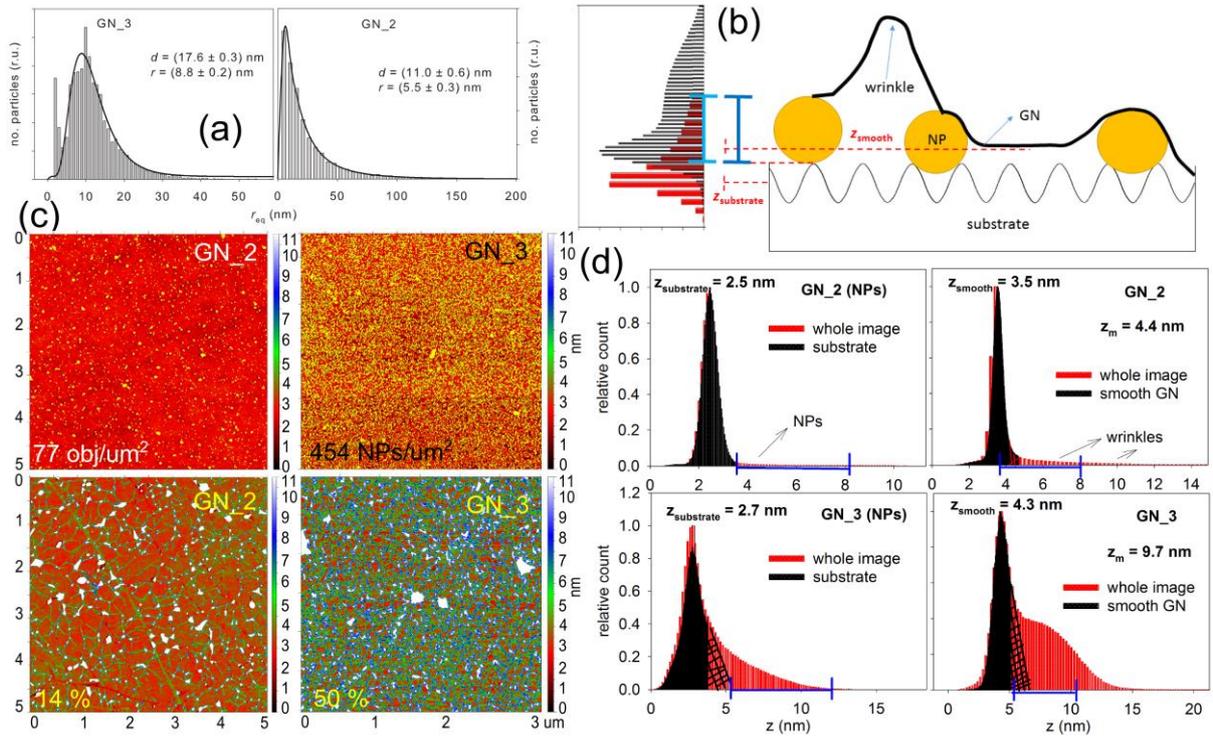

**Figure 2.** Selected statistical functions and AFM topography images for the sample with the low (GN_2) and high (GN_3) $\rho_{NP}$. (a) The histograms of the NP radius, determined from the AFM images of decorated substrates, refined with the log-normal function. Both the mean NP radius, $r$ and mean diameter, $d$ are depicted in the images. (b) Illustration of the side-cut of the GN layer on the top of the substrate decorated with the NPs, with the $z$-histograms of decorated substrates (red) and whole GN@NPs sample (black) on the left. Characteristic heights, $z_{substrate}$, $z_{smooth}$ and heights attributed to the NPs (blue line) are clearly marked. (c) AFM images of decorated substrates with masked NPs in yellow (upper images) and GN layer (lower images). The values of no. of objects/NPs per $\mu m^2$ and percentage of wrinkle surface coverage, $A_{wr}(\%)$ are depicted directly in images. (d) $z$-histograms for the substrates decorated with the NPs (left) and whole GN@NPs samples (right). The heights attributed to the NPs are marked by blue line (as in image (b)).

The statistics of the $z$ coordinate of individual pixels in the images helped to resolve whether the parts of the GN layer that are not actually forming wrinkles are in contact with the substrate or are elevated in the areas between the NPs (illustration in Figure 2b and S8-S9 in the Supplementary information file). It is the alternative of commonly used step function analysis, when the mean difference in heights between the substrate and the layer is calculated. Analysis of the $z$-histograms for decorated substrates provided values of mean height of the substrate, $z_{\text{substrate}}$, (Figure 2d), and heights attributed to the NPs. Mean distance, $d_{\text{GN\_substrate}}$, between smooth parts of the GN (with corresponding height $z_{\text{smooth}}$) and substrate, $z_{\text{substrate}}$ was determined as difference between $z_{\text{smooth}}$ and $z_{\text{substrate}}$, (Table 1). For the GN adhered at the SiO$_2$ substrate, typical thickness of the GN determined by the tapping mode AFM reaches 0.3 – 1.6 nm [58]. It was reported that large variations in the GN thickness values for similar substrates determined by different groups are result of exact setting of the scanning parameters. Error in determination of the GN thickness for the SiO$_2$ substrate was reported as 1 nm [58]. Hence we assigned the flat GN parts as delaminated from the substrate for $d_{\text{GN\_substrate}} > 2.6$ nm (the GN_5, GN_6, GN_8, GN_10 samples, Table 1).

Furthermore, wrinkling phenomena was examined in detail. It can be easily noticed that GN sheet forms three types of wrinkles - long line-shaped wrinkles (length in order of hundreds of nm-2 μm), short wrinkles (up to 500 nm) formed in between two NPs/aggregates and arrays of two, three or four short wrinkles radially escaping from the central area in which they are connected together (example Figure 1, 2c and S1-S5 in the Supplementary information file). The wrinkles were located by the threshold masking method. Using the grain analysis, the $A_{\text{wr}}$ values for individual samples were determined (Figure 3 and images visualizing the individual GN layers -Figure 2c and S1-S5 in the Supplementary information file). Comparison of the $A_{\text{wr}}$ of individual samples with other determined parameters revealed clearly that $A_{\text{wr}}$ is significantly increased for increasing roughness of the substrate. For $\sigma_p \sim 0.35 \pm 0.1$ nm, $A_{\text{wr}} \sim 30$ %, whereas $A_{\text{wr}} > 40$ % for the $\sigma_p > 0.7 \pm 0.2$ nm (see Figure S6 in the Supplementary information file). The most significant observation was linear dependence of the increasing $A_{\text{wr}}$ with increasing $\rho_{\text{NP}}$ (Figure 3).

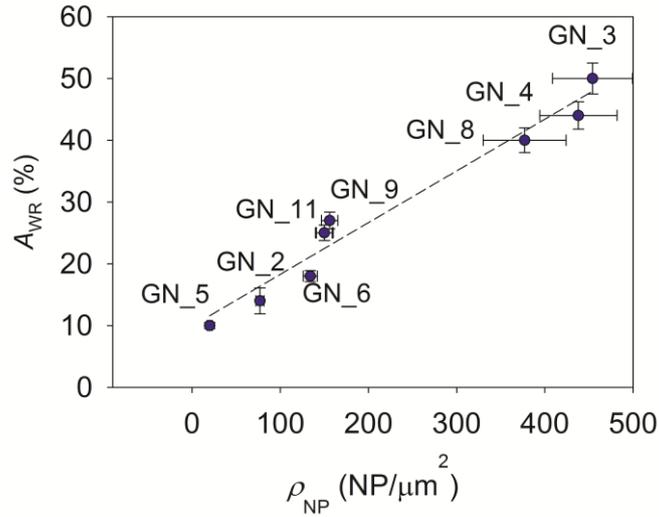

**Figure 3.** Dependence of the $A_{wr}$ on the $\rho_{NP}$ for individual samples, with the refined linear dashed line. It is obvious that the intersection of the refined linear dependence with the y axis is non-zero (~5 %), which implies there is some level of wrinkling of GN sheet for the bare $SiO_2$/Si substrate. As was reported many times in the literature, wrinkles are created also for the GN on bare substrate, presence of the NPs enhances the wrinkling.

The autocorrelation functions, HHCF of decorated substrates and whole GN layers (which is connected with the ability of surface diffusely scatters the incident light [52,56,59,60]) are summarized in the Supplementary information file (Figure S10 and S11).

2D FFT can be used as a tool to reveal the preferential direction of the orientation of wrinkles, if there is any. 2D FFT of model topography images of simple objects are depicted in Figure S14 in the Supplementary information file.

2D FFT principally enhances the sharp borders between the areas with different intensity (height in our case). If there is the sharp border in the image in some direction, it is manifested as the line in 2D FFT image (as a "ray" escaping from the central area of the 2D FFT image) in perpendicular direction. The wrinkle oriented in some direction can be viewed as such kind of border. Hence if there is the preferential orientation of wrinkles, it should be observable at the 2D FFT image of the layer. Situation for tetrahedron, as an example of the $C_{3v}$ symmetry, which is essential for the potential presence of the "pseudofield", is illustrated in Figure 4. Figure 4 a-f illustrates topography and 2D FFT of simulated tetrahedrons (created in Gwyddion) in identical and non-identical orientations (for detail explanation see caption of Figure 4). Figure 4 g-j illustrates real topography images of two samples with the corresponding 2D FFT images, with the preferential directions of orientation of wrinkles marked by red arrows (compare with Figure S14 in the Supplementary information file). The examination of selected areas with the local 3-axial orientation of wrinkles (insets in Figure 4g) showed that corresponding 2D FFT image is characteristic by the six rays diverging from the symmetric central area of the image (insets in Figure 4h).

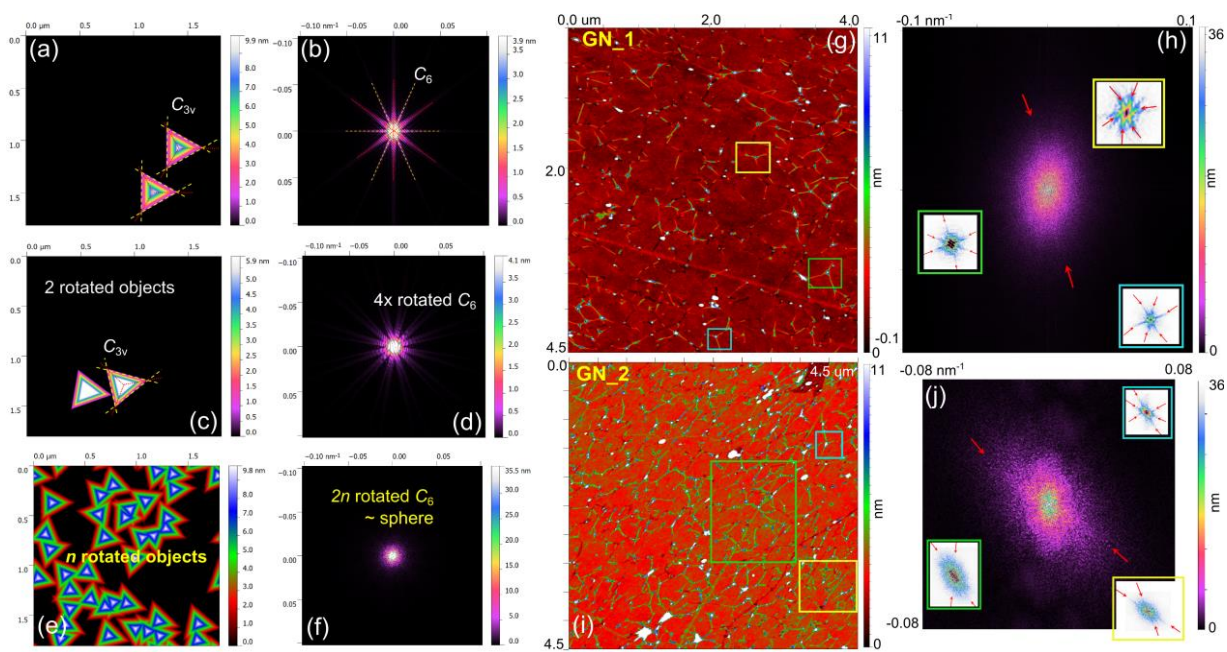

**Figure 4.** Example of the symmetry of the objects in the AFM images and relation to 2D FFT. Simulated topographical images with tetrahedrons in similar orientation (a), different orientation (c), multiple tetrahedrons with random orientation (e) and corresponding 2D FFT images (b,d,f). The yellow and red dashed lines highlight corresponding symmetry axes both in the topographical and 2D FFT images. It is obvious that $C_{3v}$ symmetry is transformed to $C_6$ symmetry, hence each topographical feature with 3-fold rotational symmetry is manifested by 6-fold rotational axis in the 2D FFT images. Global 2D FFT of the large number of randomly oriented objects with same symmetry approaches circular symmetry in (1/x, 1/y) projection. (g,i) Topographical AFM images of real samples and corresponding 2D FFT images (h, j). (h, j) illustrate that preferential orientation of wrinkles can be observed on the 2D FFT images (marked with red arrows). Selected areas of topographical images and corresponding 2D FFT are marked by yellow, green and blue rectangles. These insets represent areas of images where wrinkles are propagating only in several directions (that are labeled by arrows in the 2D FFT images). Size and the position of 2D FFT insets does not correspond to the real size and position in reciprocal space and are only illustrative.

Analysis of the phase images and correctness of determination of the NP density for the NPs hidden beneath the GN layer is summarized in the Supplementary information file (Figure S12).

It was shown that correct location of the NPs beneath the GN is possible only for the low concentration of the NPs and low $A_{wr}$. Other experiments, such as X-ray diffraction under low angles should be used instead.

**Conclusions**

We demonstrated that projected area of wrinkles created on graphene supported by the randomly distributed NPs increases linearly up to 50 % with increasing density of the NPs in a range of 20 – 470 NPs/$\mu m^2$. This result can be used for fabrication of defined level of wrinkling on the GN layer by tuning the density of objects beneath the GN layer. The wrinkling was independent on the size of the NPs, when the 6 and 10 nm large NPs were compared.

Moreover, we showed that even for such complex systems as is the GN on the top of substrates decorated with the NPs, the valuable information can be extracted with use of the AFM technique. We suggested how to easily determine delamination of the smooth parts of the GN from substrate and found preferential orientation of wrinkles, if there was any. We also showed that AFM is principally not proper tool for detection of the NPs beneath the layer. Procedure was efficient only for GN with projected area of wrinkles < 30 % and NP density lower than ~ 150 NPs/$\mu m^2$.

**Acknowledgement:**

The study was supported by the Charles University in Prague, project GA UK no. 1302313 and Grant Agency of Czech Republic (GACR), project no. 15-01953S. The γ-$Fe_2O_3$ NPs were prepared in the MULTIFUN project of the 7th framework program 7RP-262943. The Veeco Multimode V microscope was provided by the group of prof. Matolin, DSPP, Faculty of Mathematics and Physics, Charles University in Prague. We acknowledge Dr. Gorka Salas, IMDEA and ICMM Madrid and Dr. Puerto Morales, ICMM Madrid, Spain for preparation and administration of the nanoparticles for the second batch of the samples.

**Supplementary Information File for:**

**Formation of wrinkles on graphene induced by nanoparticles: atomic force microscopy study**

B. Pacakova[1,2], J. Vejpravova[1*], A. Repko[2,3], A. Mantlikova[1], M. Kalbac[4]

[1]Institute of Physics of the ASCR, v.v.i., Na Slovance 2, 182 21 Prague 8, Czech Republic.

[2]Charles University in Prague, Faculty of Mathematics and Physics, Ke Karlovu 3, 121 16 Prague 2, Czech Republic.

[3]Charles University in Prague, Faculty of Science, Albertov 6, 128 43 Prague 2, Czech Republic.

[4]J. Heyrovsky Institute of Physical Chemistry of the ASCR, v.v.i., Dolejskova 3, 182 23 Prague 8, Czech Republic.

**AFM images of decorated substrates and GN layers.**

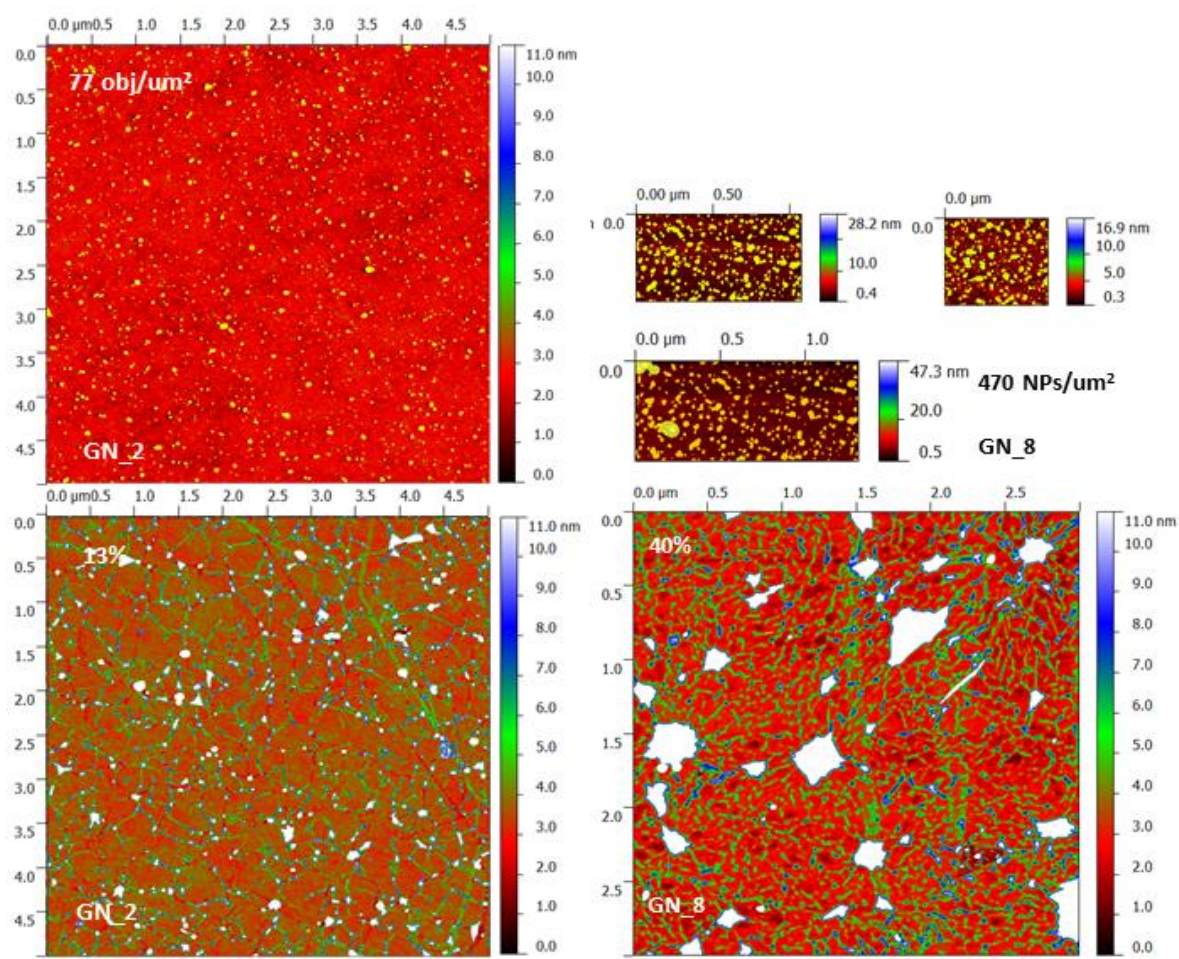

Figure S1: Top: Example of substrate decorated with the NP aggregates (GN_2) and NPs (GN_8). Bottom: Same substrates covered with the GN layer. Both the $\rho_{NP}$(/μm) and $A_{wr}$(%) are depicted in images.

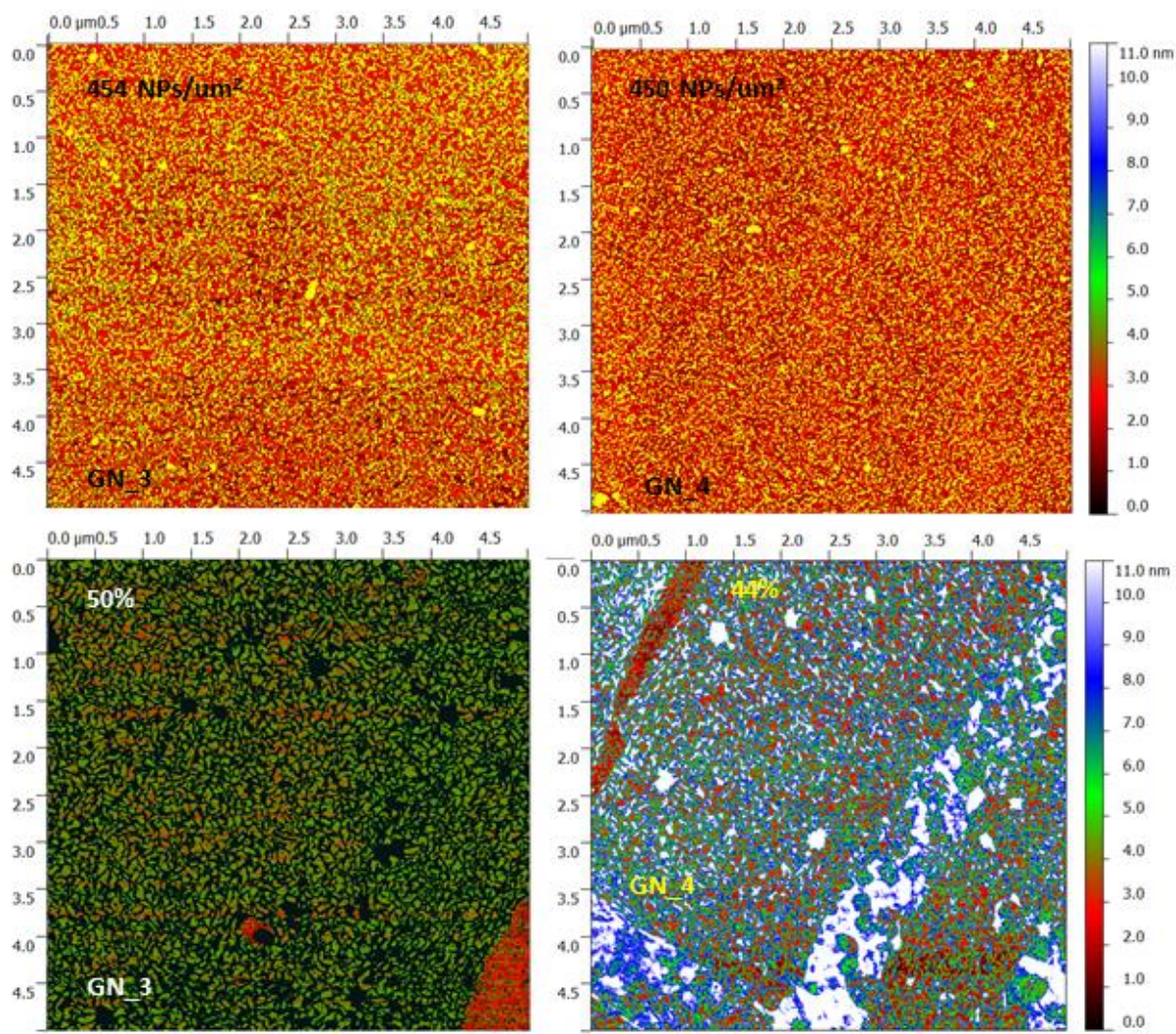

Figure S2. AFM images of the GN_3 and GN_4 samples. Top: Substrates decorated with the NPs. Bottom: Same substrates covered with the GN layer. Both the $\rho_{NP}$(/μm) and $A_{wr}$(%) are depicted in images.

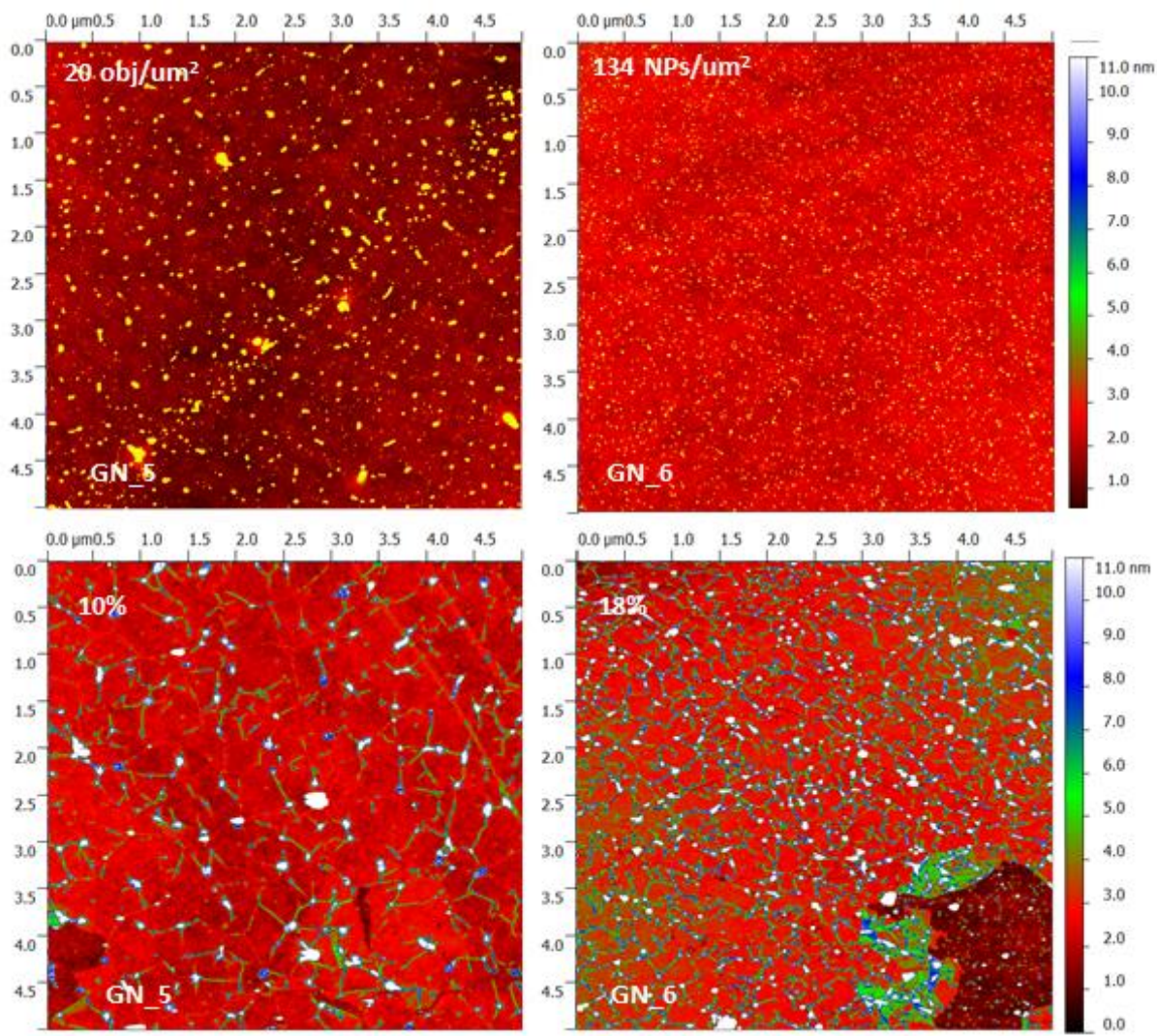

Figure S3. AFM images of the GN_5 and GN_6 samples. Top: Substrates decorated with the NPs. Bottom: Same substrates covered with the GN layer. Both the $\rho_{NP}(/\mu m)$ and $A_{wr}(\%)$ are depicted in images. It is obvious that NPs on the GN_5 sample created aggregates.

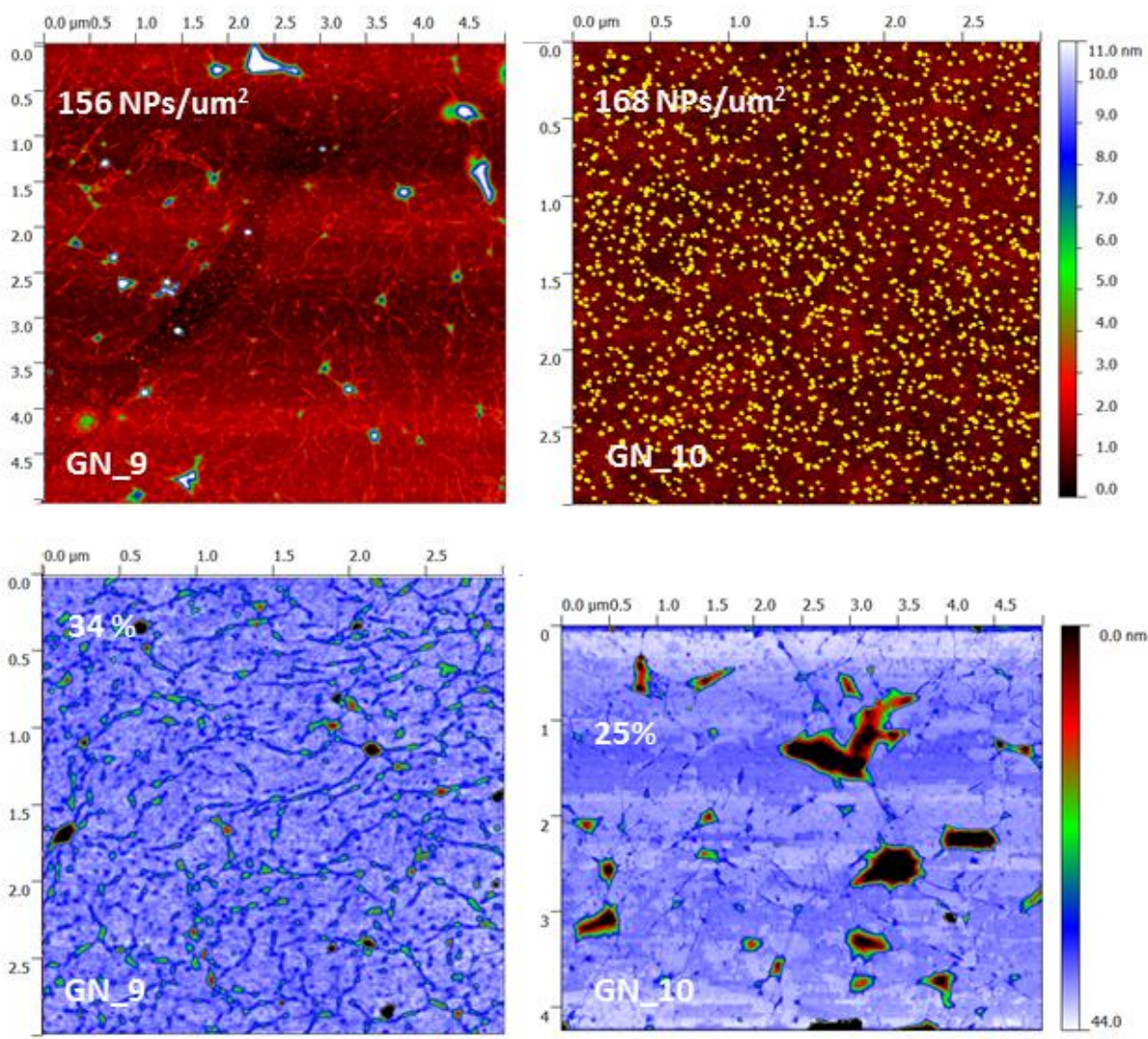

Figure S4. AFM images of the GN_9 and GN_10 samples. Top: Substrates decorated with the NPs. Bottom: Same substrates covered with the GN layer. Both the $\rho_{NP}$(/μm) and $A_{wr}$(%) are depicted in images.

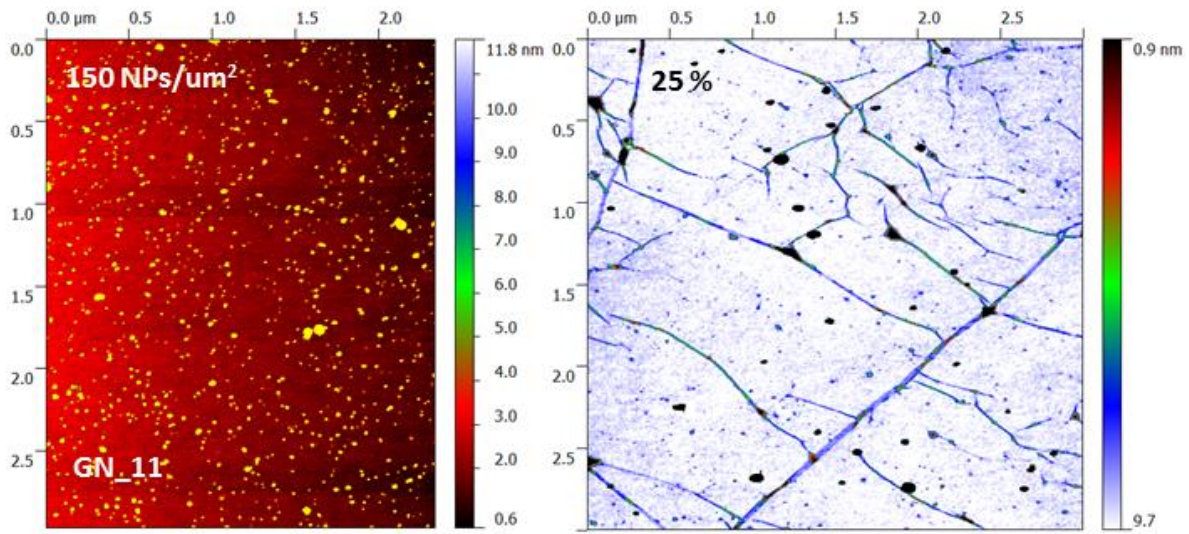

Figure S5. AFM images of the GN_11 sample. Left: Substrates decorated with the NPs. Right: Same substrates covered with the GN layer. Both the $\rho_{NP}(/\mu m)$ and $A_{wr}(\%)$ are depicted in images.

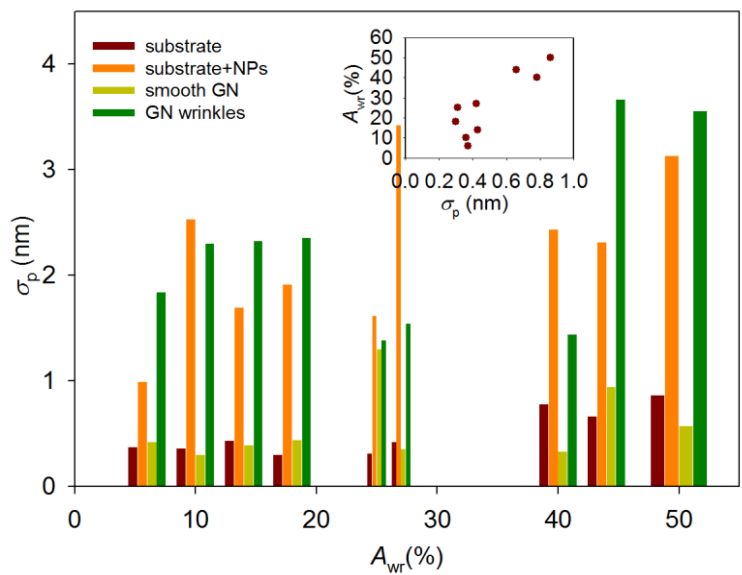

Figure S6. The dependence of $\sigma_p$ of bare substrate, decorated substrate (substrate+NPs), smooth GN and wrinkled GN on projected area of wrinkles, $A_{wr}$. Dependence of the $A_{wr}$ on the substrate $\sigma_p$.

**Histograms of equivalent disc radius, $r_{eq}$**

Following paragraph summarizes $r_{eq}$ histograms for all samples, refined with the log-normal distribution function. Obtained differences between the mean NP diameters determined from the AFM images for individual samples (Figure S8-9) are highly probably the result of different wear of the tip (NP size imaged by the probe is convolution of the real NP size and the shape of the tip), if we compare the substrates with low NP density. For the substrates with high NP density, aggregates composed of 2-3 NPs were localized, when we tried to determine the particle diameters.

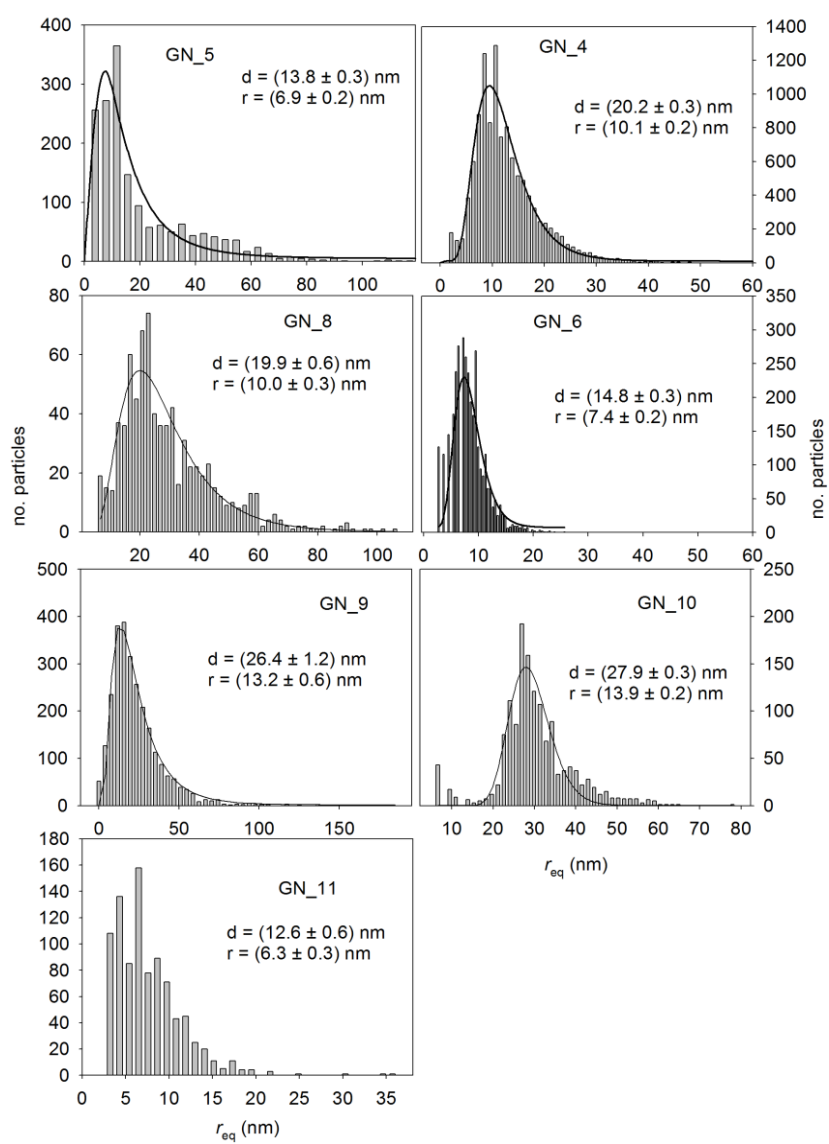

Figure S7. $r_{eq}$-histograms for the rest of the samples refined with the log-normal function. Values of mean diameter, $d$ and radius, $r$ are depicted directly in the images.

**Height, z-histograms**

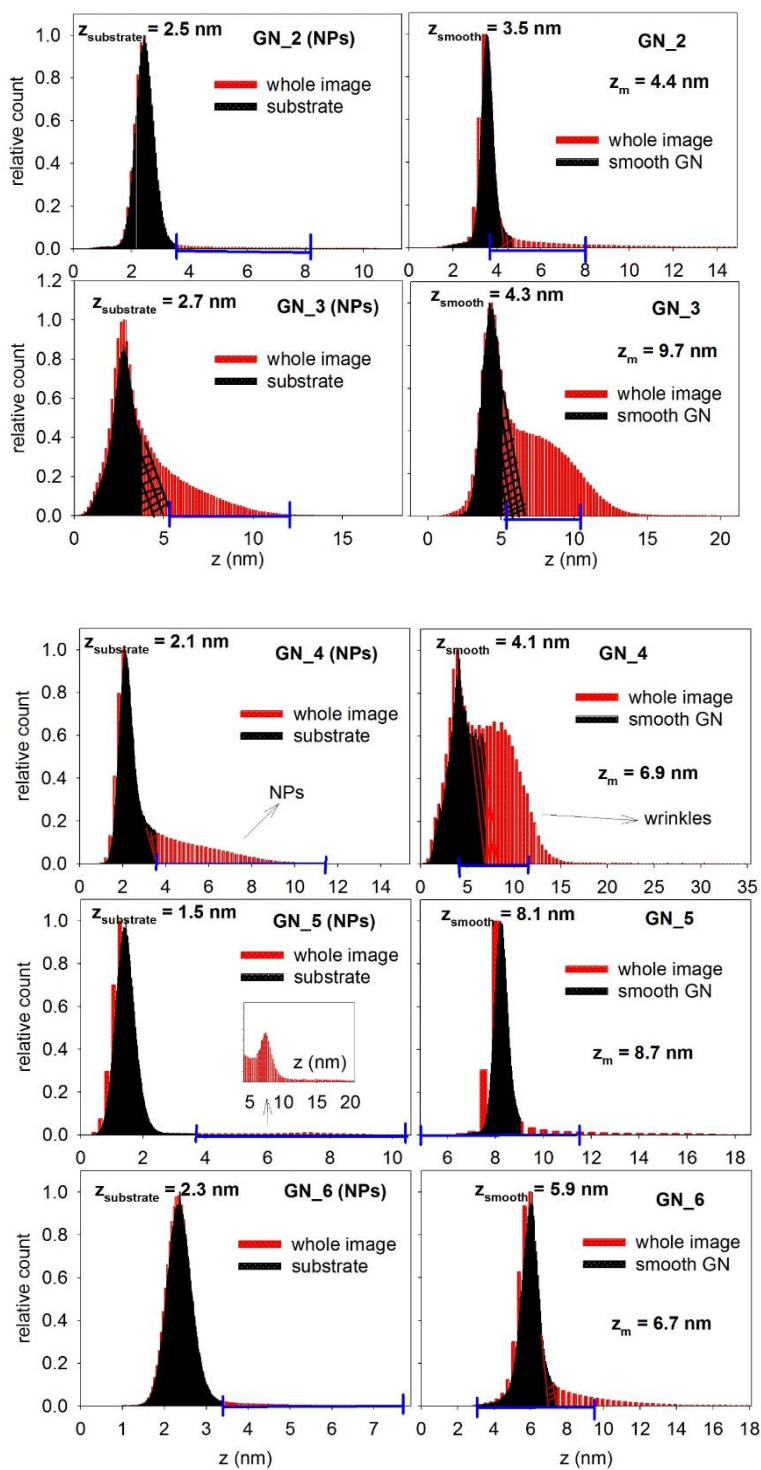

Figure S8. Example of *z*-histograms for images of decorated substrates (left) and whole GN@NPs samples for the GN_2-GN_6 samples. Blue bar represents heights attributed to the NPs.

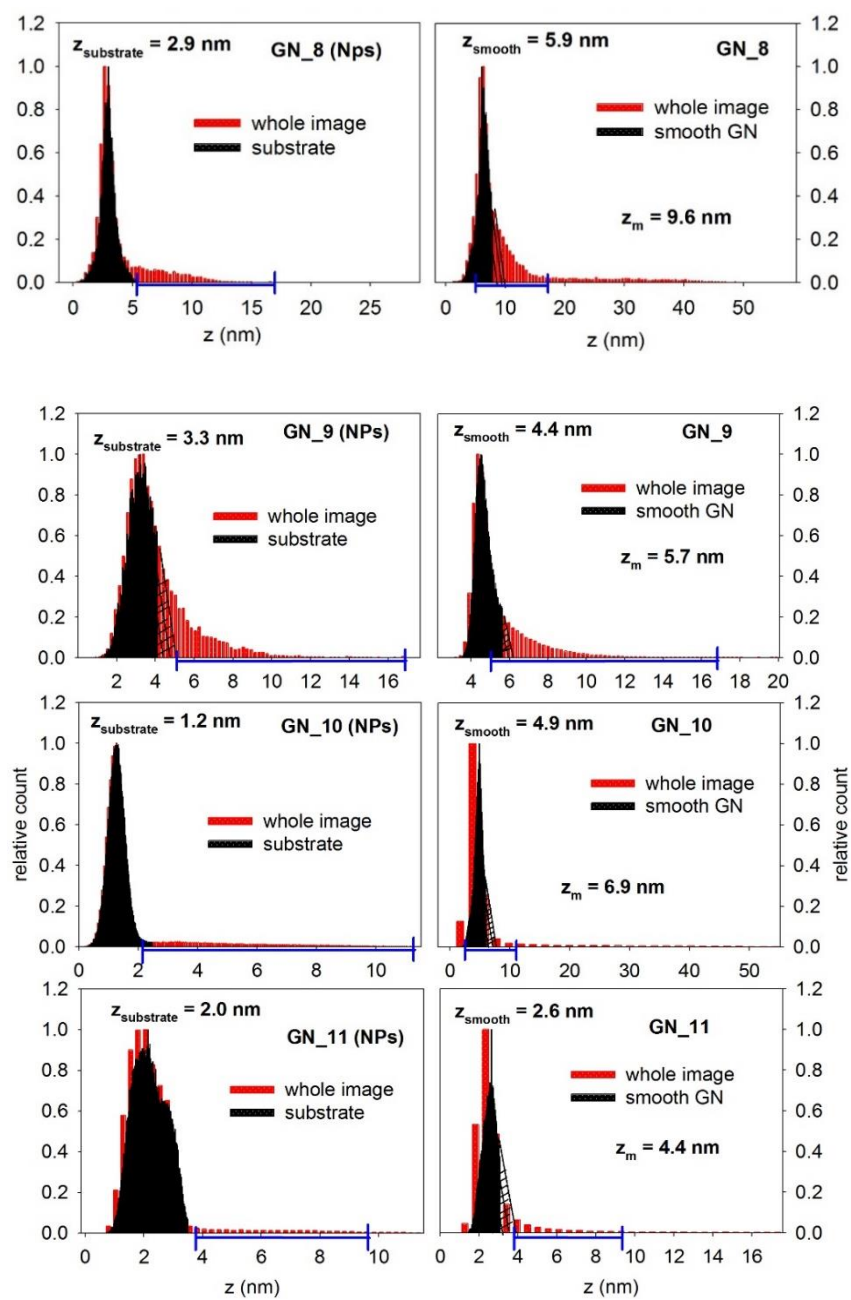

Figure S9. Example of *z*-histograms for images of decorated substrates (left) and whole GN@NPs samples for the GN_8-GN_11 samples. Blue bar represents heights attributed to the NPs.

Table S1. Table of determined parameters. NP density, $\rho_{NP}$ determined for decorated substrates, $\rho_{NP\_phase}$ determined from the phase images. Mean real and simulated interparticle distances, $d_{mean}^{real}$ and $d_{mean}^{sim}$, mean $z$ height of the wrinkled GN, $z_m$.

| Sample | $\rho_{NP}$ | $\rho_{NP\_phase}$ | $d_{mean}^{real}$ (nm) | $d_{mean}^{sim}$ (nm) | $z_m$ (nm) |
|---|---|---|---|---|---|
| GN_1 | 77 ± 5 | 62 | 146.2 ± 4.7 | 141.8 ± 5.1 | 3.5 ± 0.2 |
| GN_2 | 77 ± 5 | 76 | 139.6 ± 4.7 | 138.8 ± 5.1 | 4.4 ± 0.2 |
| GN_3 | 454 ± 45 | 430 | 54.2 ± 4.3 | 56.2 ± 4.7 | 9.7 ± 0.4 |
| GN_4 | 438 ± 44 | 460 | 54.5 ± 4.4 | 58.3 ± 4.7 | 6.9 ± 0.3 |
| GN_5 | 20 ± 3 | 23 | 267.7 ± 1.8 | 272.2 ± 1.9 | 8.7 ± 0.4 |
| GN_6 | 134 ± 8 | 65 | 104.0 ± 3.9 | 103.5 ± 4.3 | 6.7 ± 0.3 |
| GN_7 | 550 ± 55 | 745 | 40.8 ± 4.1 | 38.7 ± 4.4 | 3.7 ± 0.2 |
| GN_8 | 470 ± 47 | 123 | 58.0 ± 4.3 | 55.0 ± 4.6 | 9.6 ± 0.4 |
| GN_9 | 156 ± 9 | 80 | 96.7 ± 3.7 | 96.4 ± 4.1 | 5.7 ± 0.3 |
| GN_10 | 168 ± 10 | 525 | 92.9 ± 3.6 | 91.0 ± 4.0 | 6.9 ± 0.3 |
| GN_11 | 150 ± 9 | 731 | 106.3 ± 3.8 | 99.4 ± 4.1 | 4.4 ± 0.2 |

**Height-height autocorrelation function (HHCF)**

HHCF is one of the autocorrelation functions - the second-order statistical function describing the mutual relationship between two points at the surface. It is used for complementary description of the surfaces, as an addition to the first order statistical functions such as the height distribution or the $\sigma_q$:

$$H(\tau) = 2\sigma_q^2 \left(1 - \exp\left(\frac{-\tau^2}{T^2}\right)\right) \quad \text{(S1)},$$

where $\tau$ is the distance between the two points in the fast scan direction, and $T$ is the autocorrelation length.

Practical use of the HHCF function is connected with real roughness of the surface and diffuse scattering of incident light on such surface. It can be easily shown that for the surfaces deviating from the Gaussian like surfaces, GHCCF, significant part of the light is scattered [1–3]. Hence analysis of the HHCF function can help with predictions of the suitability of the samples for X-ray reflectivity measurements and in principle, it can be correlated with the spot size of the incident light also for other optical-based experiments, such as the Raman spectroscopy.

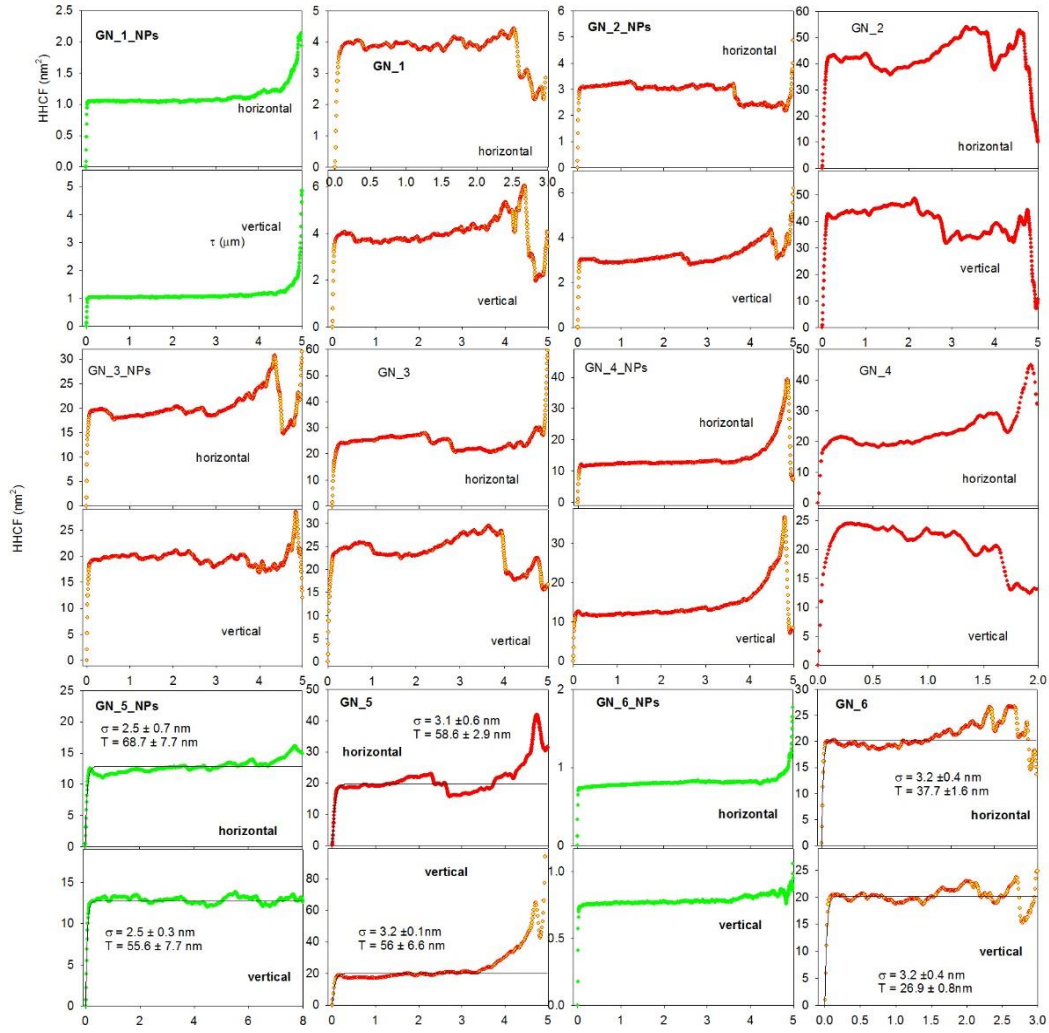

Figure S10. The HHCF of the GN_1-GN_6 samples (GN_1 sample was not involved in the further AFM study because of the corruption of the GN layer), both for the horizontal and vertical directions and decorated substrates as well as GN layers. It is obvious that even if several decorated substrates are Gaussian/Gaussian like, the statistics of the GN transferred on the top of such substrates is not preserved. The refined Gaussian HHCF are plotted in selected images (GN_5, GN_6). Results of refinement for each sample are depicted in Table S2.

Table S2. Refined parameters of the HHCF plots for individual samples in horizontal (hor) and vertical (vert) directions, both for decorated substrates (NPs) and GN layers (GN). $\sigma$ is the roughness and $T$ is the correlation length.

|        | NPs          |        |              |        | GN           |        |              |        |
|--------|--------------|--------|--------------|--------|--------------|--------|--------------|--------|
|        | hor          |        | vert         |        | hor          |        | vert         |        |
| sample | $\sigma$ (nm) | $T$ (nm) | $\sigma$ (nm) | $T$ (nm) | $\sigma$ (nm) | $T$ (nm) | $\sigma$ (nm) | $T$ (nm) |
| GN_1   | 0.7          | 20.9   | 0.7          | 20.4   | 1.4          | 28.8   | 1.4          | 26.8   |
| GN_2   | 1.2          | 19.8   | 1.3          | 19     | 4.6          | 55     | 4.4          | 47     |
| GN_3   | 3.2          | 27.3   | 3.1          | 25.4   | 3.4          | 29.9   | 3.4          | 29.9   |
| GN_4   | 2.7          | 29.8   | 2.7          | 29     | 3.5          | 39.8   | 3.1          | 34.5   |
| GN_5   | 2.5          | 68.7   | 2.5          | 55.6   | 3.1          | 58.6   | 3.2          | 56     |
| GN_6   | 0.6          | 12.4   | 0.6          | 14.9   | 3.2          | 37.7   | 3.2          | 26.9   |

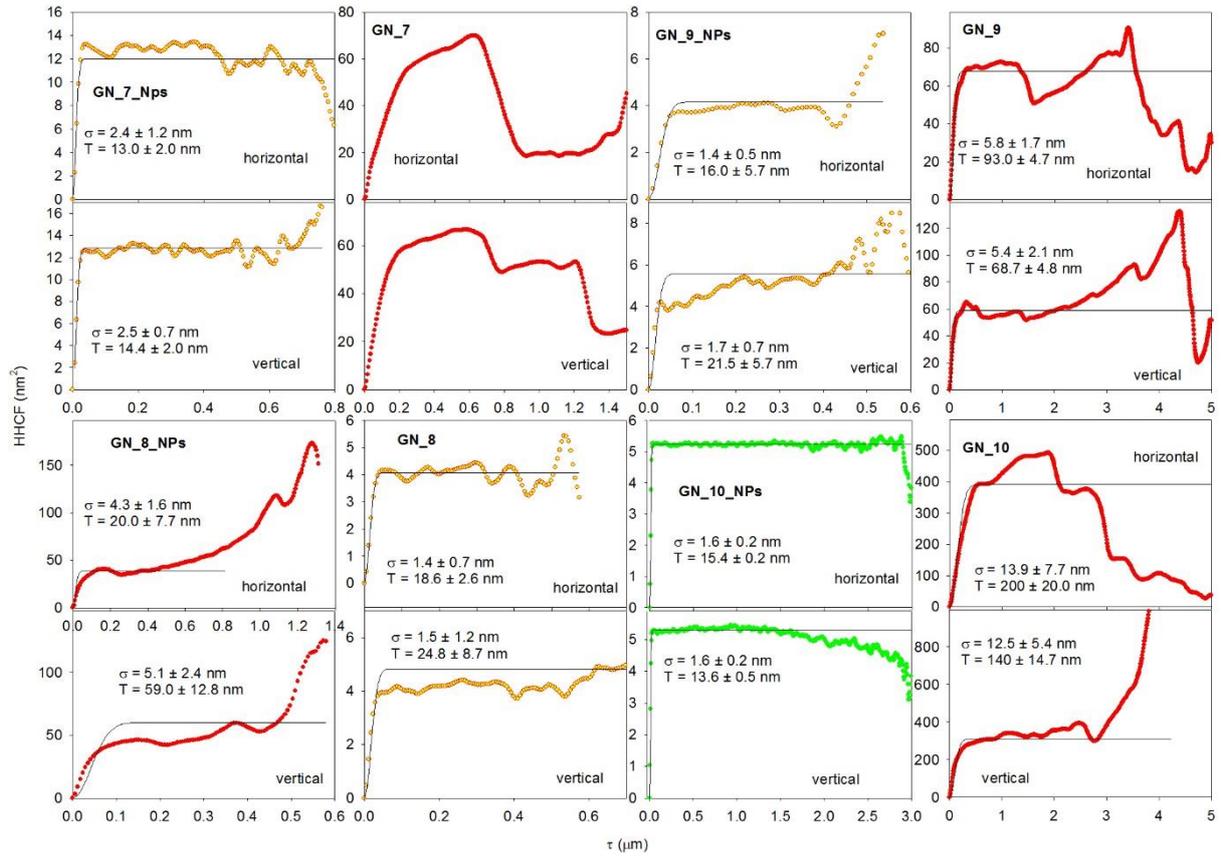

Figure S11. The HHCF of the GN_7-GN_10 samples (GN_7 sample was also not involved in the further AFM study because of the low quality of GN layer), both for the horizontal and vertical directions and decorated substrates as well as GN layers. It is obvious that even if selected decorated substrates are Gaussian/Gaussian like, the statistics of the GN transferred on the top of such substrates is not preserved.

Table S3. Refined parameters of the HHCF plots for individual samples in horizontal (hor) and vertical (vert) directions, both for decorated substrates (NPs) and GN layers (GN). $\sigma$ is the roughness and $T$ is the correlation length.

| sample | NPs | | | | GN | | | |
| --- | --- | --- | --- | --- | --- | --- | --- | --- |
| | hor | | vert | | hor | | vert | |
| | $\sigma$ (nm) | $T$ (nm) | $\sigma$ (nm) | $T$ (nm) | $\sigma$ (nm) | $T$ (nm) | $\sigma$ (nm) | $T$ (nm) |
| GN_7 | 2.44 | 13 | 2.5 | 14.4 | | | | |
| GN_8 | 4.3 | 20 | 5.1 | 59 | 1.4 | 18.6 | 1.5 | 24.8 |
| GN_9 | 1.44 | 16 | 1.66 | 21.5 | 5.8 | 93 | 5.4 | 69 |
| GN_10 | 1.6 | 15.4 | 1.6 | 13.6 | 13.9 | 200 | 12.5 | 140 |
| GN_11 | 1.55 | 22 | 1.59 | 21.2 | 2.2 | 37.4 | 2.05 | 23.5 |

**Phase images and analysis of the NP density for NPs hidden beneath the GN layer**

We have analyzed phase images of all samples. We observed that the phase image of the GN layer in contact with the substrate decorated with the NPs is characteristic by the presence of spots with elevated phase values. These spots are attributed to different measured phases – in our case the NPs hidden under the GN layer. Statistics of these spots correlates with determined NP density only in selected cases (the GN_2, GN_3, GN_4 and GN_5 samples), the values of the NP densities determined from phase images are summarized in Table S1.

However, we observed that the method gives false results for the NP densities larger than 300 NPs/um$^2$ and very large wrinkling of the layer ($A_{wr} > 30$ %), hence should be used only in the specific cases. The phase analysis of the NPs under surface is not the proper method for estimation of the NP density and bare decorated substrates should be analyzed instead.

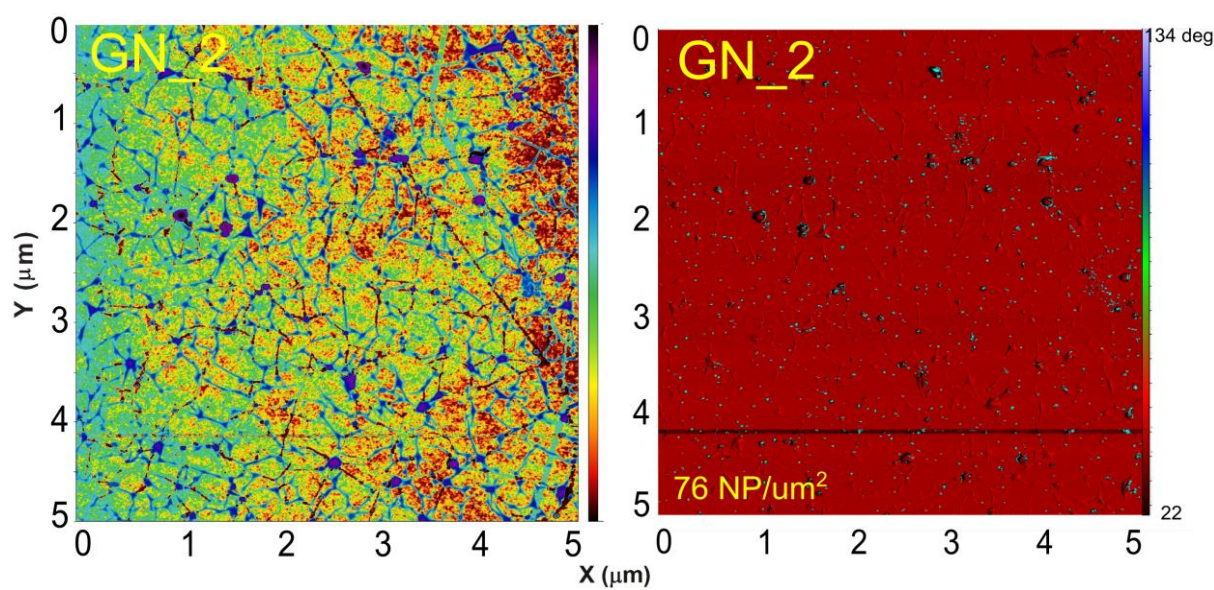

Figure S12. Height (left) and Phase (right) images of the GN_2 sample. NPs in the phase image are marked by blue.

**Orientation of wrinkles and 2D FFT**

Basic sketch of the wrinkle morphology is illustrated in Figure S13. When the wrinkle is formed, direction of strain in the flat parts of the layer is perpendicular to the main wrinkle axis (strain is represented by red arrows) [4,5] [6–8]. Hence when the wrinkles are propagating in several directions, same is valid for the strain and area with the multi-axial strain can be created in the flat parts of the layer between the wrinkles (Figure S13).

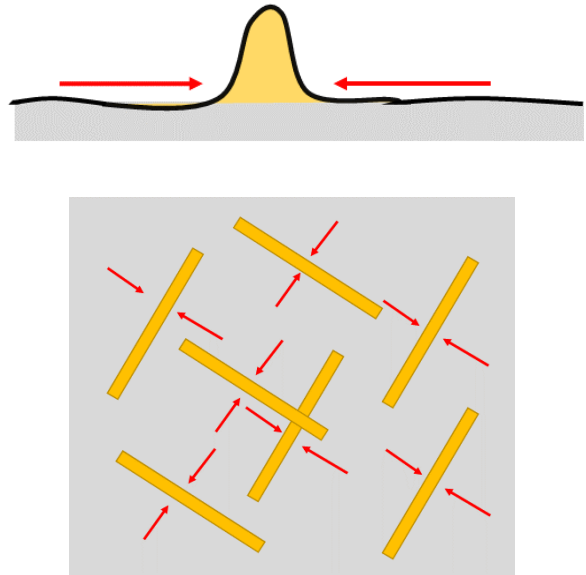

Figure S13. Illustration of the formation of wrinkles and strain in the layer (red arrows). It is obvious that if the wrinkle is formed due to the strain, the strain is perpendicular to the wrinkle propagation direction. Hence when the wrinkles are propagating in more directions, same propagation of strain, but rotated in 90 degrees can be expected.

To determine whether it is even theoretically possible to find multi-axial strain in the layer, the areas with the multi-axial propagation of wrinkles have to be found. The way how to determine preferential propagation of wrinkles is the 2-dimensional (2D) Fast Fourier transform (FFT) of the image.

2D FFT decomposes signal into its harmonic components, it is therefore useful while studying spectral frequencies present in the SPM data. It is commonly available in the SPM analysis software (Gwyddion). As an example of selected surface topographies and their 2D FFT images, see Figure S14. It is obvious that if the propagation of objects in more than one directional, 2D FFT is also manifested by propagation of spectral components in different directions. Hence 2D FFT can be an ideal tool for approval of the multi-axial propagation of GN wrinkles, and finally serves as the first assumption for existence of multiaxial strain in GN layer.

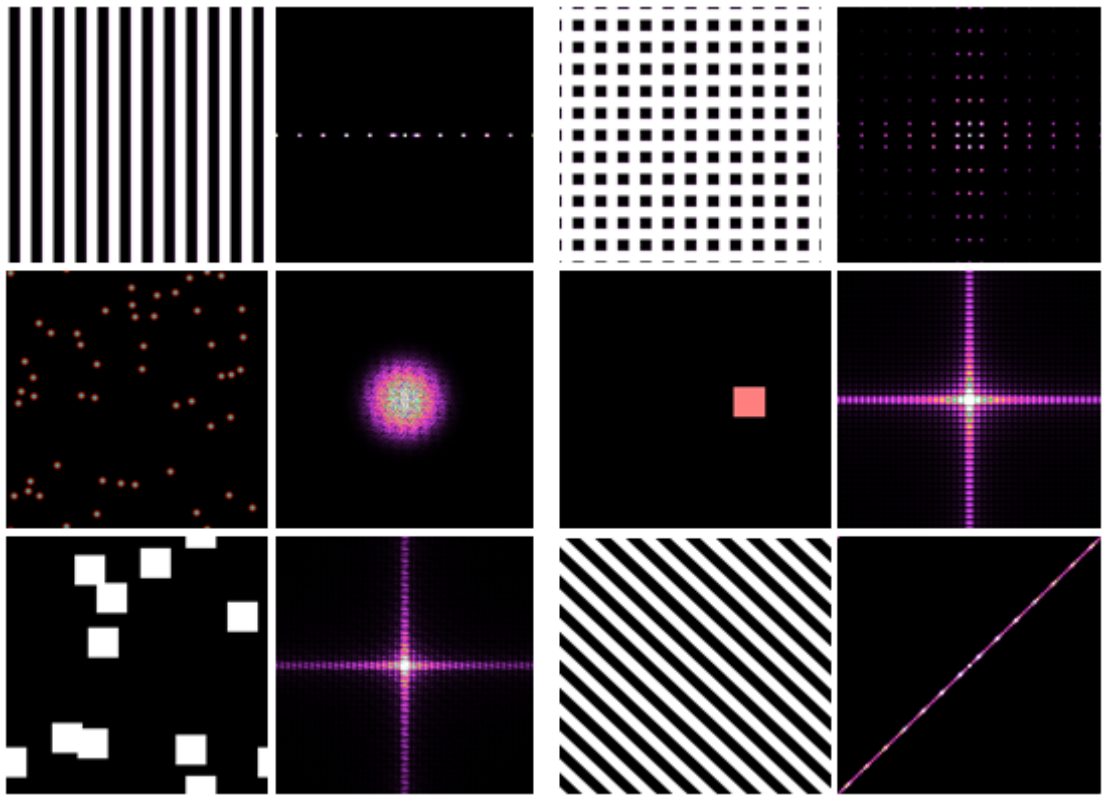

Figure S14. 2D FFT images (right columns) of simulated surfaces with stripes, spheres and boxes (left columns), respectively. Created in Gwyddion.

The images of GN layers of individual samples were subjected to 2D FFT, selected results are depicted in image S15 and S16. It is obvious that for selected samples, multi-axial propagation of objects on surface was found.

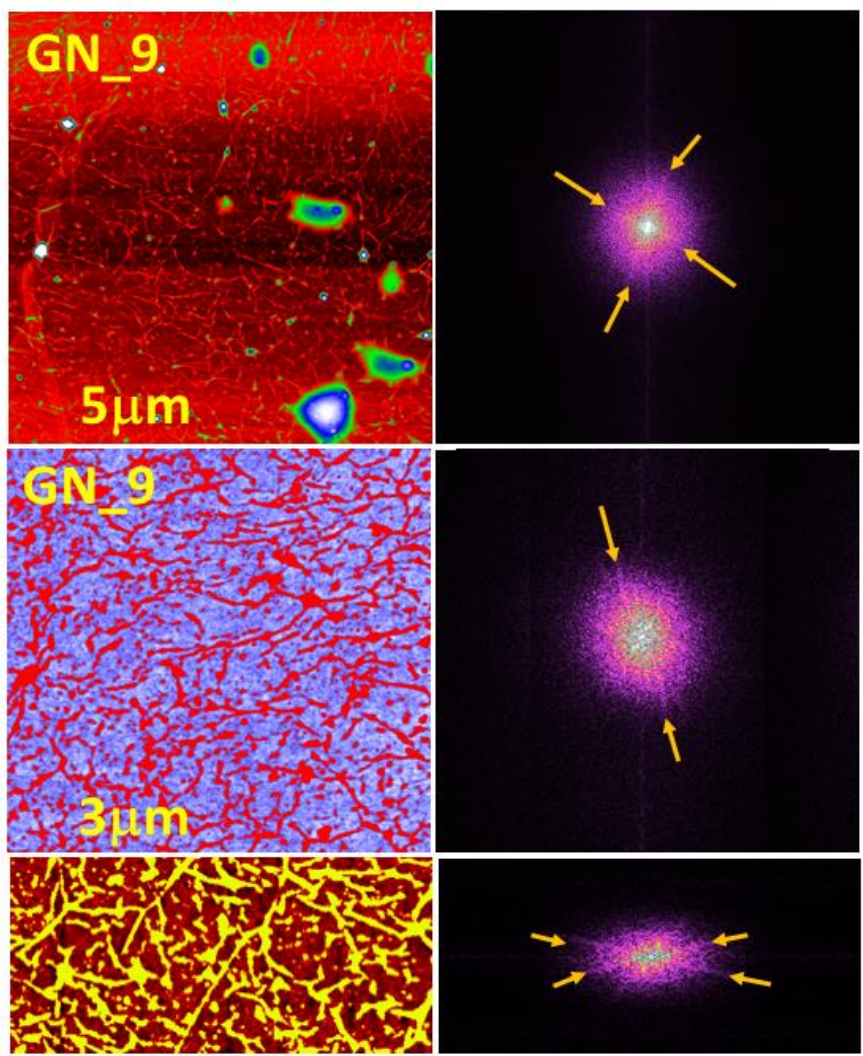

Figure S15. 2D FFT (right columns) of selected topography AFM images (left column for the GN_9 sample). Directions of preferential global propagation of wrinkles are marked by yellow arrows.

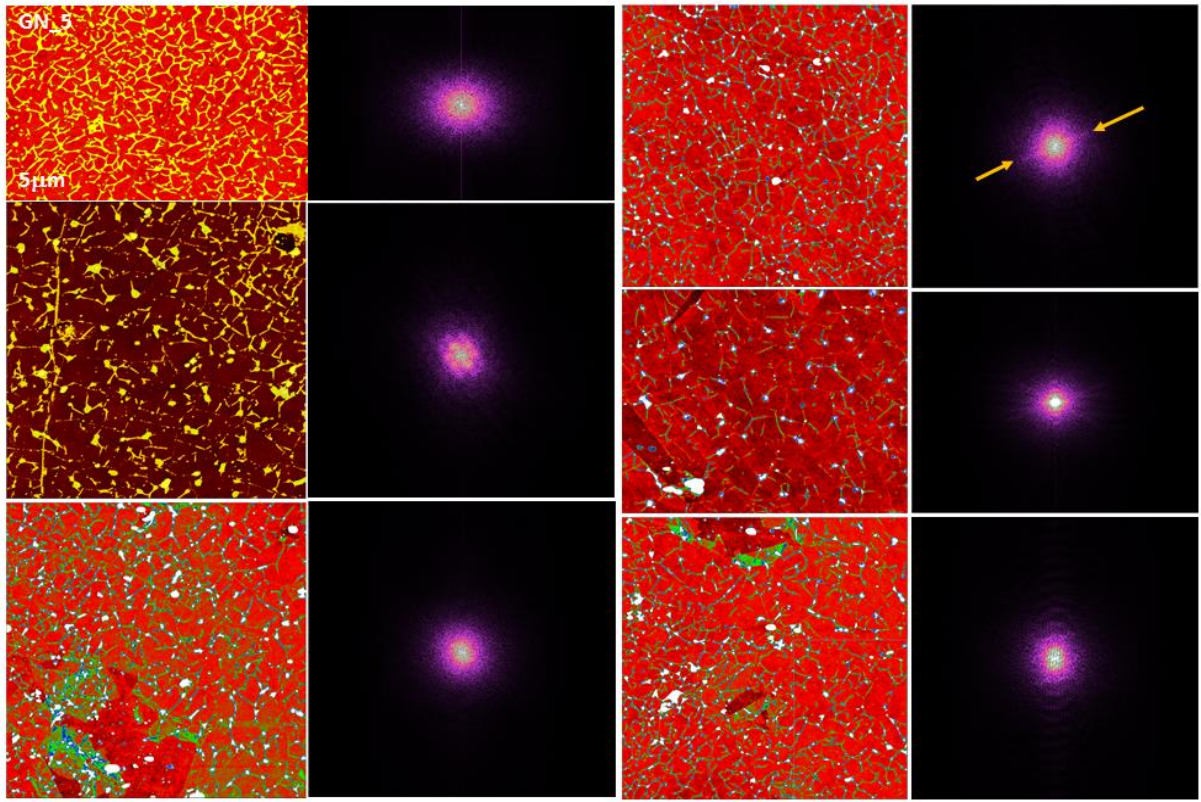

Figure S16. 2D FFT (right columns) of selected topography AFM images (left column for the GN_5 sample). Directions of preferential global propagation of objects (wrinkles) are marked by yellow arrows in the 2D FFT images.